\documentclass[conference]{IEEEtran}
\usepackage{cite}

\ifCLASSINFOpdf
   \usepackage[pdftex]{graphicx}
\else
\fi
 \usepackage{graphicx}
\usepackage{epstopdf}

\usepackage[cmex10]{amsmath}
\usepackage{tabularx}
\usepackage{listliketab}
\usepackage{lipsum}
\usepackage{multirow}
\usepackage{longtable}
\newcommand{\tabincell}[2]{\begin{tabular}{@{}#1@{}}#2\end{tabular}}

\usepackage{algorithm}
\usepackage{algorithmic}

\usepackage{array}

\usepackage{mdwmath}
\usepackage{mdwtab}

\usepackage{eqparbox}

\usepackage[tight,footnotesize]{subfigure}

\usepackage{amsmath}

\usepackage[font=footnotesize]{subfig}
\usepackage{multirow}
\usepackage{fixltx2e}

\usepackage{stfloats}

\usepackage{url}

\begin{document}
%
\title{Feasibility Study of Enabling V2X Communications by LTE-Uu Radio Interface}

\author{\IEEEauthorblockN{Ji Lianghai, Andreas Weinand, Bin Han, Hans D. Schotten}
\IEEEauthorblockA{Chair of Wireless Communication and Navigation\\
University of Kaiserslautern, Germany\\
Email: $\lbrace$ji,weinand,binhan,schotten$\rbrace$@eit.uni-kl.de}}

\maketitle

\begin{abstract}
Compared with the legacy wireless networks, the next generation of wireless network targets at different services with divergent QoS requirements, ranging from bandwidth consuming video service to moderate and low date rate machine type services, and supporting as well as strict latency requirements. One emerging new service is to exploit wireless network to improve the efficiency of vehicular traffic and public safety. However, the stringent packet end-to-end (E2E) latency and ultra-low transmission failure rates pose challenging requirements on the legacy networks. In other words, the next generation wireless network needs to support ultra-reliable low latency communications (URLLC) involving new key performance indicators (KPIs) rather than the conventional metric, such as cell throughput in the legacy systems. In this paper, a feasibility study on applying today's LTE network infrastructure and LTE-Uu air interface to provide the URLLC type of services is performed, where the communication takes place between two traffic participants (e.g., vehicle-to-vehicle and vehicle-to-pedestrian). To carry out this study, an evaluation methodology of the cellular vehicle-to-anything (V2X) communication is proposed, where packet E2E latency and successful transmission rate are considered as the key performance indicators (KPIs). Then, we describe the simulation assumptions for the evaluation. Based on them, simulation results are depicted that demonstrate the performance of the LTE network in fulfilling new URLLC requirements. Moreover, sensitivity analysis is also conducted regarding how to further improve system performance, in order to enable new emerging URLLC services.  
\end{abstract}


%
\IEEEpeerreviewmaketitle

\section{Introduction}
Current 4G wireless communication network offers good connectivity for most of the time. However, in areas with poor coverage, under excessive interference or in a situation where network resources are overloaded, the reliability of wireless link is not guaranteed \cite{D62}. For the year beyond 2020, a new generation of wireless systems will be designed to offer a solution with a high degree of reliability and availability for commercial wireless systems, in terms of rate, latency or another Quality of Service (QoS) parameter \cite{D11}. In order to meet the corresponding demand for new service types, such as vehicular communications to improve traffic safety and moving vehicular networks, intensive research work is performed to design the next generation wireless communication system that is able to fulfill the new requirements of URLLC and corresponding services.\\ 
Compared with conventional KPIs, e.g. overall cell throughput and per link data rate, which were representatively used to evaluate legacy wireless systems, URLLC has different technical goals as \cite{URLLC}:
\begin{itemize}
\item	10 to 100 times higher number of connected devices: URLLC through the reliable service decomposition providing mechanisms for an increased number of users.
\item	10 times longer battery life for low power devices: URLLC providing beneficial in disaster applications when using low power devices.
\item	5 times reduced E2E latency: URLLC providing reduced end-to-end latency in real-time applications.
\end{itemize}
Note that above figures are obtained with a baseline comparison of 4G network.\\
As a typical scenario for URLLC application, the communication taken place between two nearby vehilces has strict requirements on latency and transmission reliability. The information exchange between any traffic participants can contribute to the cooperative driving and therefore improve the traffic efficiency and safety. To provide solutions to the considered scenario, a lot of work in literature focus on the exploitation of a direct vehicle-to-anything (V2X) communication, where the data packet is directly transmitted from the transmitter to the receiver without going through the network infrastructure. For instance, the 3rd Generation Partnership Project (3GPP) proposes to use a PC5 interface to facilitate the direct communication between the two ends of a V2X communication \cite{23303}. Moreover, European Telecommunication Standards Institute (ETSI) Intelligent Transport Systems (ITS) propose to apply IEEE 802.11p protocol as the air interface for the direct V2X communication \cite{80211}. With the proposed direct V2X communication, the transmission latency can be efficiently reduced, since network infrastructure is not involved in the data transmission.\\
It is worth noticing that, when a direct V2X communication takes place among a set of users, it is performed by a point-to-multi-point multicast approach. In this way, all relevant traffic participants can get the information of the transmitter. Therefore, a base station (BS) or a Roadside Unit (RSU) can also receive the multicasted message\cite{22185}. A RSU a computing device locates on the roadside and provides connectivity support to passing vehicles. Moreover, the RSU can either be connected to the BS with a backhaul link or even act as a base station \cite{23285}. Therefore, the network can obtain the transmitted packets, and further transmit these packets to the target receivers through the LTE-Uu air interface which is used for the connection between the eNodeB and mobile users in LTE network. With this approach, a diversity gain can be obtained on top of the proposed direct V2X communication. Therefore, an evaluation on the feasibility of using LTE-Uu interface to offer V2X service should be carried out, which is the main target of this work.\\
In this work, we carry out an evaluation of applying LTE network and LTE-Uu interface to enable V2X communication \cite{D11}. The performance can also be considered as a baseline to evaluate any new relevant technologies in future.\\ 
This paper is organized in the following structure. Firstly, the system model and methodology to evaluate packet E2E latency and successful packet transmission rate are introduced in Sect.~\ref{SM}. Afterward, in Sect.~\ref{SA}, the relevant simulation assumptions are stated. In Sect.~\ref{SR}, we demonstrate simulation results, where the performance of both unicast and multicast transmission modes in downlink are inspected on. Further, sensitive analysis is also conducted in Sect.~\ref{SR}, where the future work to provide the support for V2X service by the cellular network is highlighted. Finally, we draw the conclusion of our work in Sect.~\ref{con}.
\section{System Model and Simulation Methodology}\label{SM}
In order to improve traffic efficiency and to aid the driver to avoid the occurrence of an accident, cooperative intelligent traffic systems (C-ITS) \cite{ITS}, which rely on the timely and reliable exchange of information of traffic participants, can be exploited. This refers to the collection of the state information of other vehicles (e.g., position, velocity and acceleration) through wireless communications and also the information exchange between vulnerable road users (VRUs) (e.g., pedestrians) and vehicles. The main challenge here corresponds to the strict requirement on the low packet transmission latency and high reliability.\\
\begin{figure}[!t]
\centering
\includegraphics[width=3.5in]{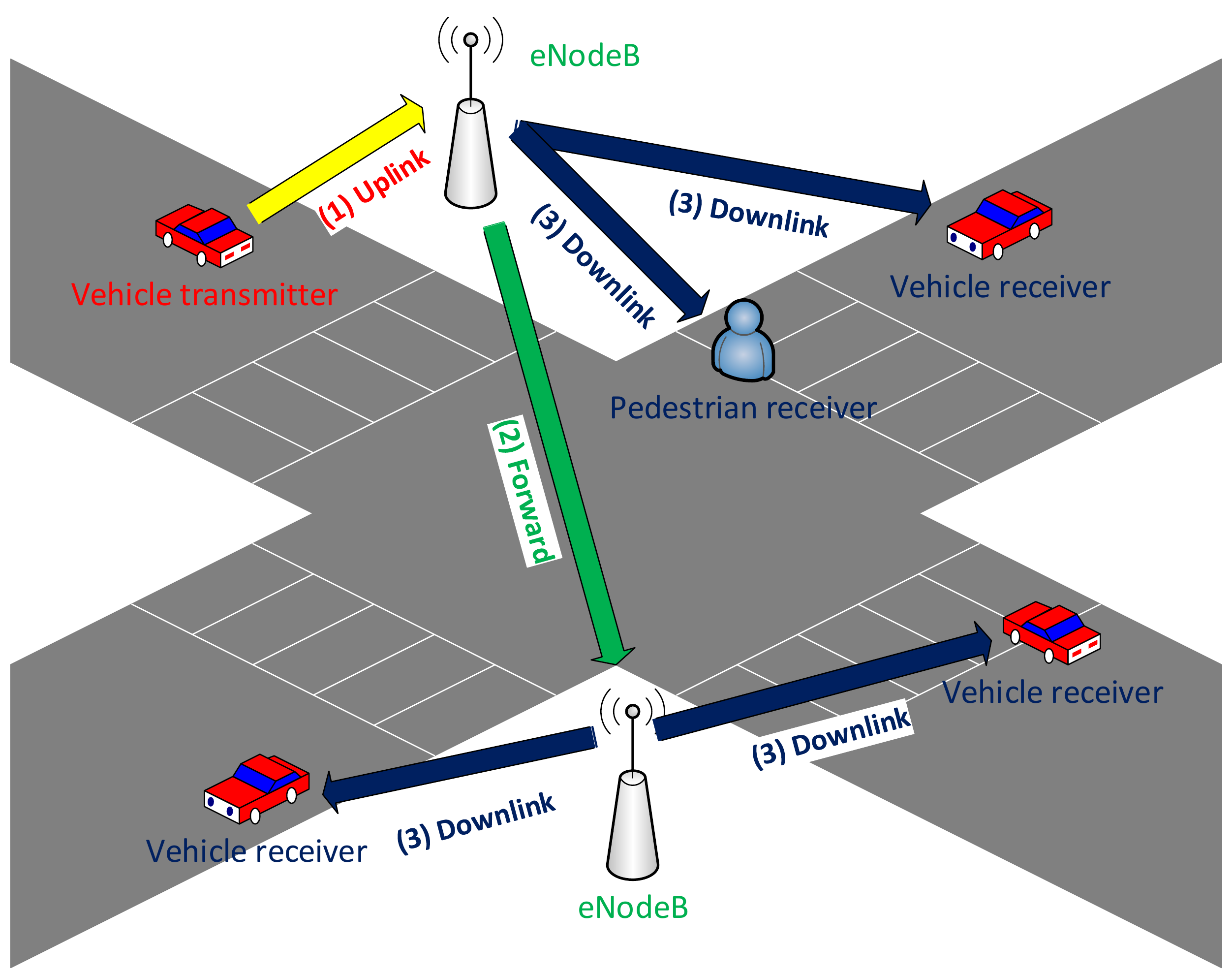}
\centering
\caption{V2X transmission scenario in LTE network}
\label{v2v}
\end{figure}
The above information exchange refers to a communication process, as:
\begin{enumerate}
\item The transmitter vehicle transmits its user data packet to the serving eNodeB by LTE-Uu in uplink.
\item The eNodeB forwards the received packet to the serving eNodeBs of the receivers.
\item Packets are further transmitted to the receivers from their serving eNodeBs.
\end{enumerate}
In Fig.~\ref{v2v}, we showcase the this process. From this figure, user-plane (UP) E2E latency can be divided into three parts from the perspective of logical packet position in the LTE network, as follows:
\begin{itemize}
\item Uplink latency - represents the time difference between the generation of one packet at the transmitter and its successful arrival at the serving eNodeB.
\item Propagation latency between eNodeBs - represents the packet propagation time between the serving eNodeB of the transmitter and the serving eNodeB of one receiver.
\item Downlink latency - represents the time difference between the packet arrival at the serving eNodeB of the receiver and its successful arrival at the receiver.
\end{itemize}
Please note, the UP E2E latency is the one way transmission time of a packet between the transmitter and the successful arrival of this packet at the receiver.
\begin{figure}[!t]
\centering
\includegraphics[width=3.5in]{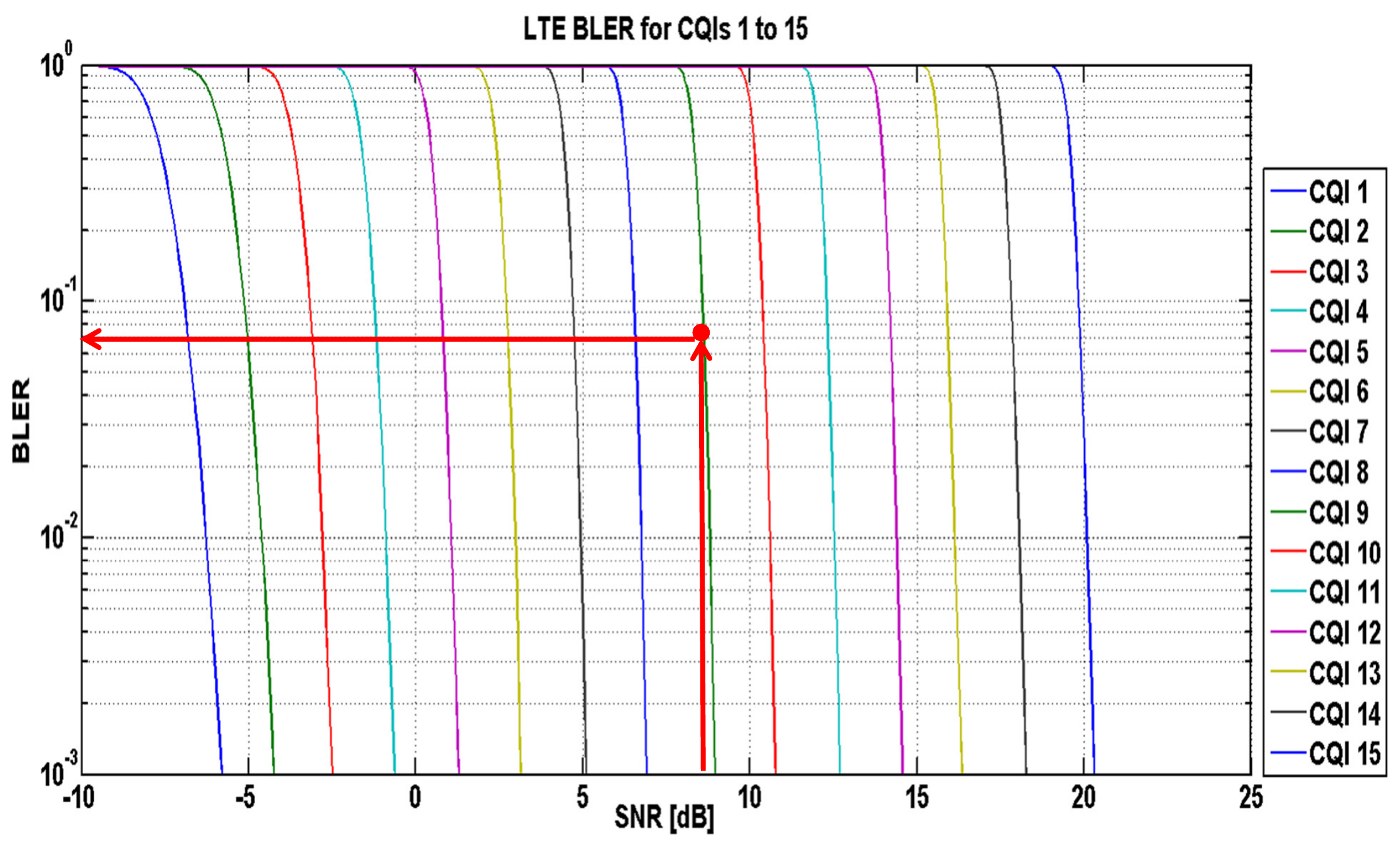}
\centering
\caption{Mapping from SNR to block error rate}
\label{BLER}
\end{figure}
\subsection{Uplink latency}
The uplink latency can be evaluated by Alg.~\ref{alg1} where each packet is traced from the generation until its successful receiving by the corresponding serving eNodeB. Note that the latency here is inspected in the UP, therefore the control plane (CP) functions (e.g., connection setup, mobility control and radio access procedure) are not considered in this work. Besides, to illustrate the step~\ref{ACMuplink} and \ref{BLERuplink} in Alg.~\ref{alg1}, an example is given in Fig.~\ref{BLER} to show the mapping from signal-to-noise-ratio (SNR) to both adaptive coding and modulation scheme (ACM) and block error rate (BLER) in LTE network \cite{wien}. In order to fit the transmission strategy with radio link quality, the transmitter selects the coding and modulation scheme which provides the maximal date rate with a BLER less than $10\%$. Thus, the BLER of one radio link can be derived from its SNR.
\subsection{Propagation latency between eNodeBs}
Packet propagation through the serving eNodeB of one transmitter to the eNodeB of one receiver may involve the behavior of the core network(CN). However, we assume an enhanced X2 interface for the connection between any two nearby eNodeBs and therefore a low propagation latency and a high reliability of the data packet transmission among different eNodeBs can be achieved. With this assumption, the previous mentioned E2E latency components originate mainly from the LTE-Uu air interface in uplink and downlink. Moreover, in case the transmitter and all the target receivers are served by the same eNodeB, the data packet does not need to be transmitted to another eNodeB.
\subsection{Downlink latency}
Once a packet is available at the eNodeB of the receiver, the eNodeB schedules certain time and frequency resource for the packet transmission to the receiver. With the LTE-Uu air interface, two transmission modes can be utilized in the downlink transmission.\\
The first option corresponds to a unicast transmission procedure where the packet is delivered to the receivers in a point-to-point manner. In this approach, if there are $N$ users to receive the packet, then the eNodeB needs to transmit the packet for $N$ times. To be noticed, this unicast mode allows the eNodeB to apply different coding and modulation schemes for the different receivers, as long as their channel state information (CSI) are available at the eNodeB. Additionally, once the packet is decoded at the receiver, an acknowledgment(ACK)/non-acknowledgment(NACK) feedback message is transmitted back to the eNodeB. And the eNodeB can decide whether a retransmission is to be triggered or not, based on this feedback message. The downlink packet transmission latency with a unicast transmission mode from the eNodeB to its receivers can be evaluated by Alg.~\ref{alg2}.\\
Another alternative refers to a multicast transmission in a point-to-multi-point manner. As mentioned in Sect.~\ref{SM}, the V2X communication refers to a process where the information of one vehicle is distributed to other vehicles in its proximity. Therefore, the multicast transmission mode supported by LTE-Uu interface can be exploited to provide this service. In this manner, the eNodeB only needs to multicast the packet to $N$ receivers for once. In order to successfully deliver the packet to all the target receivers, the eNodeB needs to apply a coding and modulation scheme taking into account of the worst case receiver. The worst case receiver, in this sense, points to the receiver which experiences the worst radio propagation channel from the eNodeB. Moreover, one another characteristic of the LTE multicast service is the absence of the feedback message. It means that the receivers will not feed an ACK/NACK message back to the transmitter, no matter whether the reception is successful or not. As an efficient approach to improve the transmission reliability, if there are enough available time and frequency resource, the eNodeB can repeat its transmission with the same content. In this case, if a packet is not successfully received, the receiver performs the maximal ratio combining (MRC) process, based on the received replica(s). To evaluate the E2E latency for the multicast transmission mode in downlink, Alg.~\ref{alg3} describes the methodology.
\begin{algorithm}
\caption{Evaluation of uplink latency}
\label{alg1}
\begin{algorithmic}[1]
\STATE A packet is generated at one vehicle. In a period of 100 ms, packets generation time among different vehicles has a uniform distribution. 
\STATE	Perform transport block cyclic redundancy check (CRC) attachment and code block segmentation on each packet.
\STATE	Decide coding and modulation scheme w.r.t. SINR value of each transmitter.\label{ACMuplink}
\STATE	BLER is derived from the SINR value of each transmitter.\label{BLERuplink}
\STATE	Round robin scheduler is used to determine how many frequency resource blocks and TTIs are allocated to each uplink packet. 
\STATE	Uplink packet starts to be transmitted to the serving eNodeB.
\STATE	If a packet is not received error free, w.r.t. BLER of step~\ref{BLERuplink}, we start ARQ retransmission and inspect on whether ARQ retransmission is possible and successful. 
\STATE	Once a packet is successfully received by serving eNodeB, the time instance of when this packet is received is recorded. If the packet transmission is not successful, the packet delay is considered as infinity.
\end{algorithmic}
\end{algorithm}	
\begin{algorithm}
\caption{Evaluation of downlink latency}
\label{alg2}
\begin{algorithmic}[1]
\STATE Only packets successful received by eNodeBs in the uplink will be transmitted in downlink.
\STATE A packet arrives at eNodeB, packet arrived time is decided by the uplink and the propagation latency between the eNodeBs.
\STATE Perform transport block CRC attachment and code block segmentation on each packet.
\STATE	Decide coding and modulation scheme w.r.t. SINR value of each receiver.
\STATE	BLER is derived from SINR value of each receiver.\label{BLERdownlink}
\STATE	eNodeB allocates the time and frequency resource to the most recently received packets. In case if multiple packets are ready to be transmitted simultaneously, round robin scheduler is used to determine how many frequency resource blocks are allocated to each downlink packet. \label{scheduleDown}
\STATE	Downlink packet starts to be transmitted to receiver.
\STATE	If a packet is not received correctly, w.r.t. BLER of step~\ref{BLERdownlink} , we start ARQ retransmission and inspect on whether ARQ retransmission is possible and successful. 
\STATE	Once a packet is successfully received by receiver, the time instance is recorded. If the packet transmission is not successful, the packet delay is considered as infinity.
\end{algorithmic}
\end{algorithm}
\section{simulation assumptions}\label{SA}
In this work, a dense urban scenario \cite{D61} is used here for simulation purpose and the detailed simulation assumptions of this scenario can be found in \cite{D61}. Due to the page limited, only the important information related to V2X communications is highlighted in the following.
\begin{figure}[b]
\includegraphics[width=3.3in]{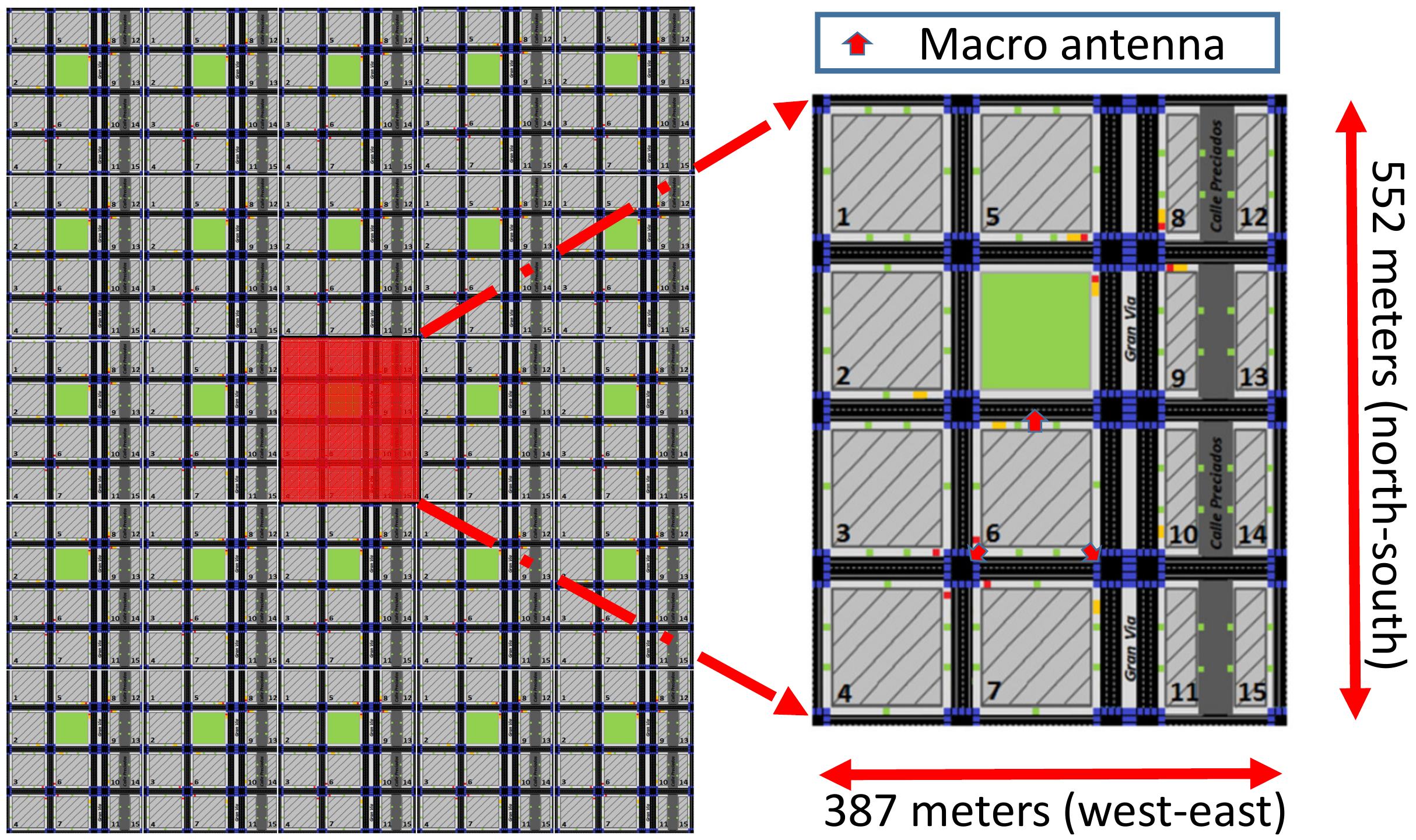}
\flushleft
\caption{environment model}
\label{model}
\end{figure}
\begin{table}
\caption{UP latency parameters}
\label{LP}
\begin{center}
\begin{tabular}{|c|c|}
\hline Description & Value\\ \hline
 UE processing delay & 1 ms\\ \hline
 Frame alignment& 0.5ms\\ \hline
 TTI for UL/DL packets &Packet specific \\ \hline
 HARQ retransmission & 7 ms\\ \hline
 eNB processing delay & 1 ms \\ \hline
 Packet exchange between eNBs & 1 ms \\ \hline
\end{tabular}
\end{center}
\end{table}
\subsection{Environment model}
In order to characterize the real world environment, a Madrid-grid environment model \cite{D61} is used, as shown in Fig.~\ref{model}. As can be seen, each Madrid-grid (i.e., colored as red) includes 15 buildings and one park.
\subsection{Deployment model}
An operating frequency of 800 MHz and a dedicated bandwidth for V2V and V2P communications are assumed here. Therefore, no traditional cellular users reuse the same resource with V2X communications. A macro base station with three sectors is deployed on the roof of the central building, as shown in Fig.~\ref{model}. At the beginning, an LTE FDD mode with 10 MHz for uplink and 10 MHz for downlink is used. The bandwidth resource of both uplink and downlink will be increased step by step to check the performance gain in Sect.~\ref{SR}. An antenna is installed on top of each vehicle and a constant transmission power of 24 dBm is used for vehicle transmitter.
\subsection{Traffic model}
A packet with an average size of 212 bytes is generated by one vehicle with 10 Hz periodicity \cite{packet} and this packet should be delivered to all vehicles and VRUs located within a radius of 200 meters. Moreover, traffic participants are deployed with a density of 1000 users/$\text{km}^2$ on the street in the central Madrid-grid, as colored by red in Fig.~\ref{model}.\\
\begin{algorithm}[t]
\caption{Evaluation of multicast latency}
\label{alg3}
\begin{algorithmic}[1]
\STATE Only packets successfully received by eNodeBs in uplink will be transmitted in downlink.
\STATE A packet arrives at eNodeB, packet arrived time is decided by the uplink and the propagation latency between eNodeBs.
\STATE Perform transport block CRC attachment and code block segmentation on each packet.
\STATE	Decide coding and modulation scheme based on the network configuration, taking into account of the state of the worst case receiver.
\STATE	The eNodeB allocates its time and frequency resource to the most recently received packets. In case if multiple packets are ready to be transmitted simultaneously, round robin scheduler is used to determine how many frequency resource blocks are allocated to each downlink packet. \label{scheduleMulti}
\STATE  Within the allocated resource, BS multicasts the packet. In case with extra available resource, the packet transmission will be repeated, in order to fully utilize the available resource. \label{repeat}
\STATE	BLER is derived from the SINR value of each receiver. Based on the BLER value, whether a packet transmission is successful or not can be inspected. \label{BLERmulti}
\STATE In case a packet reception is failed and a repetition of this transmission is applied, an HARQ process referring to the chase combining is carried out at the receiver. The HARQ process will introduce a new effective SINR value and correspondingly a new BLER. With this new BLER value, the step~\ref{BLERmulti} is repeated.
\STATE If a packet is successfully received, the reception time instance will be recorded. If a packet is not successfully received in the allocated time resource, the packet will be discarded by the eNodeB and considered with an infinity delay.
\end{algorithmic}
\end{algorithm}
\subsection{Mobility model}
Since users are deployed in an urban scenario, a maximal velocity of 50 km/h is assumed, which corresponds to the maximal velocity allowance in the most European cities.
\subsection{User-plane latency parameters}
In Tab.~\ref{LP}, the relevant UP latency parameters for our evaluation are listed \cite{912}. To be noticed here, the minimum delay between the end of a packet and the start of a retransmission is 7 ms in LTE FDD mode \cite{book} and this value is used in our work when ARQ retransmission is required. Besides, even though transmitted packets have a fixed number of 212 bytes, the number of transmission time intervals (TTIs) required for one packet transmission with the unicast transmission mode is packet specific, since it depends both on the spectrum efficiency of the applied modulation and coding scheme (MCS) and frequency resource blocks allocated to this packet transmission.\\
\subsection{Key performance indicators}
The relevant KPIs for our simulation are itemized and explained in the following:
\begin{itemize}
\item Packet E2E latency: this latency value is calculated as the time difference between packet generation time at the transmitter and the successful receiving time of this packet at the receiver. 
\item Average latency: mean value of packet E2E latency w.r.t. successful transmission. Note that the unsuccessful transmission is not taken into account here, since it has a latency value of infinity.
\item Successful packet transmission rate: the ratio between the number of the successfully received packets and the number of packets which should be delivered.
\end{itemize}
\section{Simulation results and sensitivity analysis}\label{SR}
\subsection{Simulation Results}
\begin{figure}[t]
\centering
\includegraphics[width=3.4in]{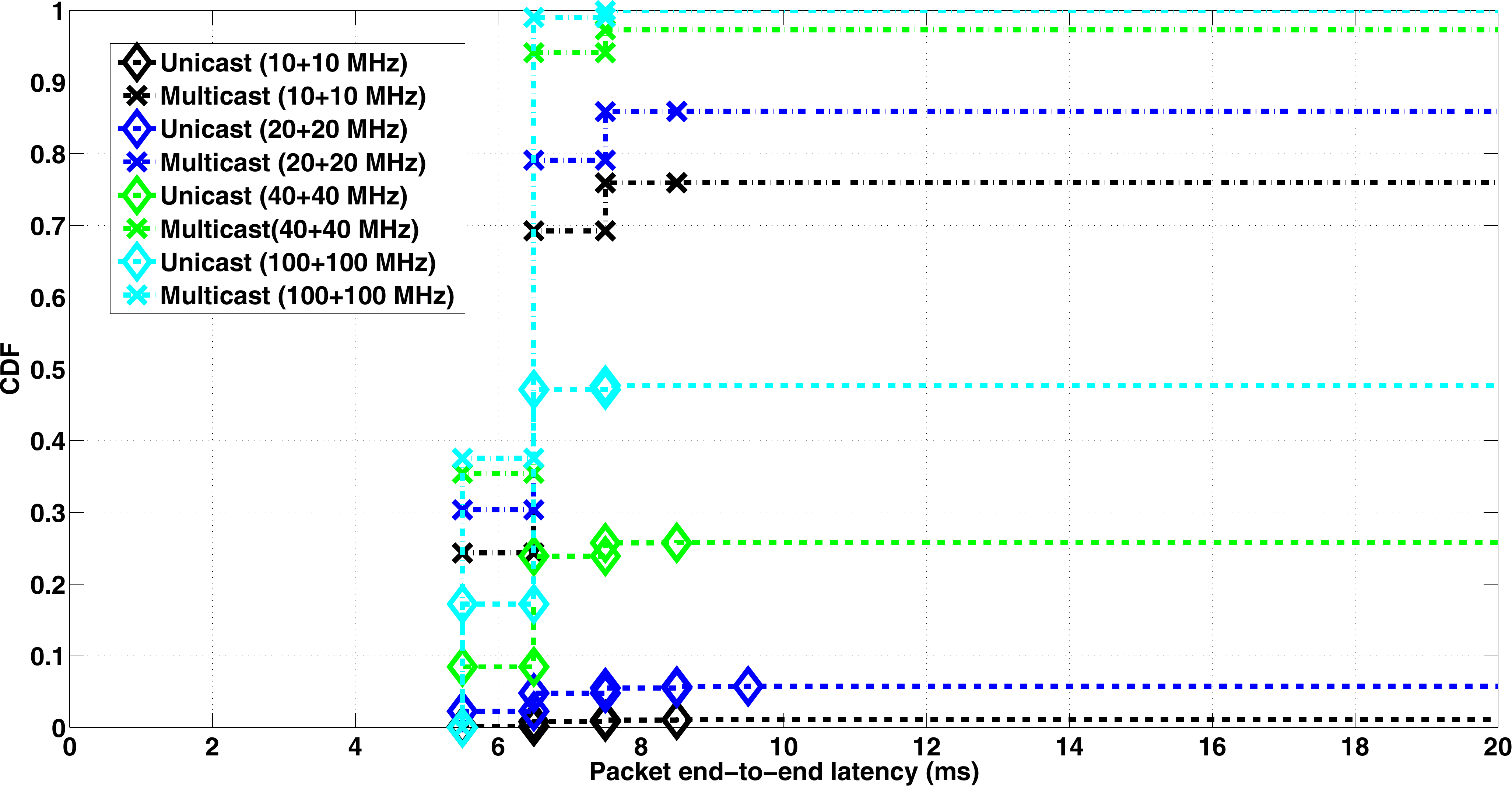}
\caption{Performance comparison between Unicast and Multicast (Spectral efficiency of multicast = 0.877 bits/Hz)}
\label{cdfplot}
\centering
\end{figure}

\begin{figure}
\includegraphics[width=3.4in]{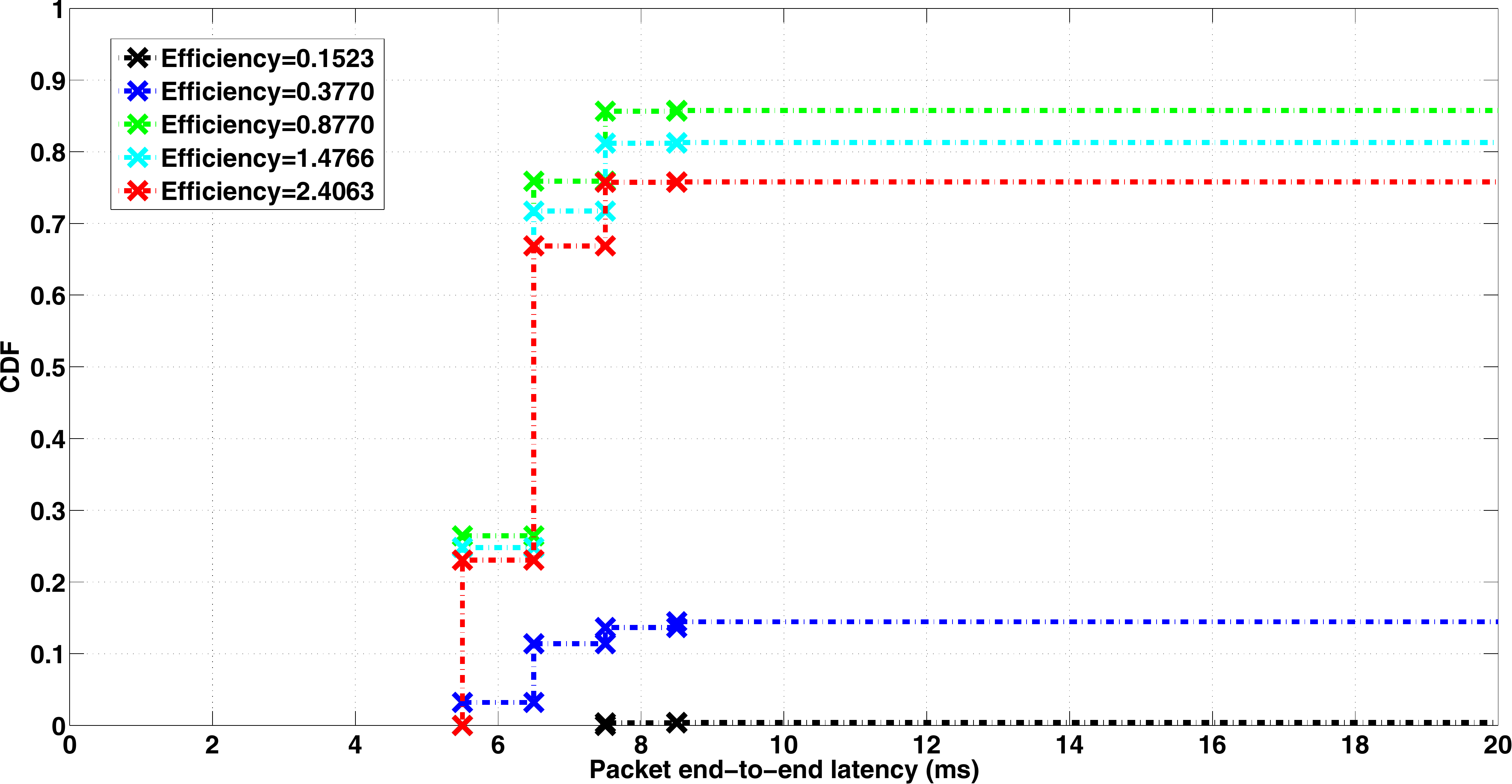}
\centering
\caption{CDF plot of packet E2E latency by multicasting}
\label{cqi}
\end{figure}

Fig.~\ref{cdfplot} shows the cumulative distribution function (CDF) plot of the packet E2E latency w.r.t. different bandwidth dedicated to V2X communications. This plot demonstrates the system performance with a series of bandwidth, ranging from 10 MHz~/~10 MHz to 100 MHz~/~100 MHz for uplink/downlink. Additionally, the average latency and the successful packet transmission ratio are also listed in  Tab.~\ref{AL}. It can be seen from both Fig.~\ref{cdfplot} and Tab.~\ref{AL}, the unicast transmission mode in downlink requires a large bandwidth to provide the V2X communication with the considered user density. And this is the reason why a significant performance improvement can be achieved with an increased system bandwidth. As a comparison, the multicast transmission mode, which applies an MCS with a spectral efficiency of 0.877 bits/Hz, obviously outperforms the unicast transmission mode, by using the same amount of bandwidth resource. To be noticed, if a receiver fails in the reception of the transmitted packet, it tries to receive the second repetition of the same packet. This is the reason why steps occurred in the curves for the multicast transmission case.\\
In Fig.~\ref{cqi}, the system performance regarding the multicast transmission in downlink are shown, where different MCSs with different efficiency are applied. The plot shows us the importance of applying an appropriate MCS to the system. In case a MCS with a lower efficiency is applied, more resource is required for the downlink transmission. Therefore, due to the lack of resource, the two MCSs with the efficiency of 0.1523 bits/Hz and 0.377 bits/Hz perform worse than the MCS with an efficiency of 0.877 bits/Hz. However, from another perspective, an MCS with higher efficiency requires a good channel state and therefore it is not robust for the receivers with low SINR values. Therefore, we can see from the figure that the two MCSs with the efficiency of 1.4766 bits/Hz and 2.4063 bits/Hz are not so robust and many receivers fail in the packet reception. And the MCS with an efficiency of 0.877 bits/Hz performs better than them. Thus, when the network configures the transmission efficiency of the multicast transmission, a compromise between robustness and efficiency needs to be achieved, based on the real-time system load and the radio condition of the receivers. 

\subsection{Sensitivity Analysis}
Based on the observation of the simulation results, the bottlenecks and challenges of applying LTE-Uu interface to enable the new emerging service are analyzed and highlighted in this subsection.
\begin{itemize}
\item	Large number of users in downlink: since one packet generated at transmitter should be received by multiple receivers in the proximity of the transmitter, this number of receivers is quite large when a big radius of target communication area is under inspection. Therefore, though the packets can be successfully received by eNodeBs in uplink, they may not always be successfully delivered to receivers by using unicast in downlink due to the huge traffic load in downlink. As a solution, the multicast transmission mode can mitigate this problem efficiently.
\item	Coding and modulation scheme: the coding and modulation scheme is critical since it determines the BLER of a packet transmission w.r.t. a specified SINR value. In unicast transmission mode, coding and modulation scheme is adapted to reach a target BLER value below $10\%$, and that means the BLER of a specific link is in the scale between $0$ to $10\%$ and the unsuccessful received packets should be retransmitted with HARQ scheme if extra resource are available. Since the minimum delay between the end of a packet transmission and the start of a retransmission is 7 ms in LTE FDD mode, in order to guarantee a certain value of packet E2E latency, the number of packet retransmissions is required to be below a certain level. Thus, using more robust coding and modulation scheme can help to decrease the number of retransmissions.
\item	Frame structure: due to the protocol design of LTE, the minimum delay of 7 ms in the retransmission procedure does not efficiently support the latency-critical services. Thus, a more flexible frame structure developed for URLLC is necessary in order to decrease the minimum delay for the packet retransmission.
\item	Scheduler: in the current LTE network, scheduling algorithms aim at serving users with the consideration of overall cell throughput, which is not the main challenge for URLLC type of services. Therefore, in order to meet the requirement of URLLC type of services (e.g., introduce a small latency value and meanwhile offer a high successful packet transmission rate), latency-dependent scheduling algorithms should be inspected and developed.
\item	Physical layer technology: physical layer technology design is highly relevant for the spectrum efficiency and therefore determines the number of required resource for one packet transmission. Therefore, transmission technology offering higher spectrum efficiency can simultaneously provide URLLC type of services to a larger number of devices.
\item	Unicast transmission through eNodeB: it is to be noticed that, the simulation results shown previously does not mean that the multicast transmission should fully replace the unicast in downlink. In case when the eNodeB only needs to transmit the packet to a small amount of receivers, unicast can outperform the multicast transmission, due to the adaptive coding and modulation scheme applied for each receiver and the existence of the feedback message. A typical scenario is the V2X communication at late night, where there are few outdoor traffic participants.
\item Diversity gain: a diversity gain can be achieved if both the LTE-Uu interface and direct V2X air interface are used simultaneously, since one packet will be transmitted to the receiver through different links.
\end{itemize} 
\section{Conclusion}\label{con}
In this work, we propose the evaluation methodology and assumptions to evaluate the impact of V2X communication on the legacy network. LTE-Uu interface is considered and modeled to derive the system performance. The simulation results demonstrate the feasibility of applying the LTE network and LTE-Uu interface to provide the considered services. In order to fulfill the requirement on the strict E2E latency and high successful packet transmission rate for V2X communications, a multicast in downlink can be exploited, together with a network controlled D2D communication, to achieve a diversity gain. This evaluation work also reflects the fact that the goals of URLLC in next generation wireless systems cannot be achieved by simply changing a parameter in the system design, but rather to develop a crafted architecture, taking all technologies into account from the physical layer to other higher layers.
\begin{table}
\caption{Average latency and successful transmission rate}
\label{AL}
\begin{center}
\begin{tabular}{|c|c|c|c|c|}
\hline \tabincell{c}{Bandwidth\\(UL+DL)} &\tabincell{c}{Mean latency\\ by unicast} &\tabincell{c}{Successful\\rate by\\ unicast} &\tabincell{c}{Mean Latency\\by multicast} &\tabincell{c}{Successful\\rate by\\multicast}\\ \hline
  10+10&6.649 ms&1.07$\%$ &6.269 ms&75.9$\%$\\ \hline
 20+20& 6.323 ms&5.75$\%$ &6.227 ms&85.91$\%$\\ \hline
 40+40 &6.246 ms& 25.75$\%$ &6.168 ms&97.26$\%$\\ \hline
   100+100& 6.151 ms&47.68$\%$ &6.133 ms&99.91$\%$\\ \hline
\end{tabular}
\end{center}
\end{table}
\section{Acknowledgment}
A part of this work has been supported by the Federal Ministry of Education and Research of the Federal Republic of Germany (BMBF) in the framework of the project 5G-NetMobil. The authors would like to acknowledge the contributions of their colleagues, although the authors alone are responsible for the content of the paper which does not necessarily represent the project.

%
%



%

\end{document}